\documentclass[finallayout,an,fleqn]{w-art}
\usepackage{times}
\usepackage{w-thm}
\theoremstyle{plain}

\theoremstyle{definition}

\usepackage[]{graphicx}
\chardef\bslash=`\\ 

\def\lesssim{\mathrel{\hbox{\rlap{\hbox{\lower4pt\hbox{$\sim$}}}\hbox{$<$}}}}
\def\gtrsim{\mathrel{\hbox{\rlap{\hbox{\lower4pt\hbox{$\sim$}}}\hbox{$>$}}}} 
\def\arcsec{\hbox{$~\!\!^{\prime\prime}$}}

\def\deg{\hbox{$^\circ$}}

\def\Md{M$_{\odot}$}

\def\HII{H\,{\sc ii}}

\begin{document}
\DOIsuffix{theDOIsuffix}
\Volume{324}
\Issue{S1}
\Copyrightissue{S1}
\Month{01}
\Year{2003}
\pagespan{1}{}
\Receiveddate{3 March 2003}
\keywords{Galaxy:center-galaxies: individual (Sagittarius)-techniques: interferometer}
\subjclass[pacs]{}



\title[Discovery of Sgr A$^*$]{The Discovery of Sgr A$^*$}


\author[Sh. First Author]{W.M. Goss\footnote{Corresponding
     author: e-mail: {\sf mgoss@nrao.edu}, Phone: +01\,505\,835\,7267,
     Fax: +01\,505\,835\,7027}\inst{1}} \address[\inst{1}]{National Radio Astronomy
Observatory}
\author[Sh. Second Author]{Robert L. Brown\inst{1,2}}
\address[\inst{2}]{National Astronomy and Ionospheric Center}
\author[Sh. Third Author]{K.Y. Lo\inst{1}}
\begin{abstract}
  The galactic center compact radio source Sgr A$^*$ was discovered on 13 and
15 February 1974 by Bruce Balick and Robert L. Brown using the Green Bank 35 km
radio link interferometer (Balick \& Brown 1974). We discuss other observations of this source
in the years 1965-1985. Early VLBI observations 
 are  described. The name \emph{Sgr A$^*$} was first used by
Robert L. Brown (1982) and has become the accepted name for the compact source
at the center of the Milky Way.
\end{abstract}
\maketitle                   


\renewcommand{\leftmark} {W.M. Goss et al. : Discovery of Sgr A$^*$}



\section{Introduction}

The discovery of Sagittarius A as a radio source coincident with the
center of the galaxy has been discussed by Goss \& McGee (1996). Almost
20 years after the recognition that the center of the galaxy could
be associated with Sgr A by Piddington \& Minnett (1951) and McGee \& Bolton
(1954) , Sgr A$^*$ was discovered in February 1974 by Bruce Balick and
Robert L. Brown in Green Bank, West Virginia. This discovery is certainly
one of the more important galactic radio astronomy discoveries 
of the 1970's and has had wide ramifications during the last 30 years. As 
an example, the recognition that the radio source Sgr A$^*$ is the dim
radio source  associated with a $2.6 \times 10^6$ \Md\/ black hole
has represented a fundamental advance in our understanding of the nuclei of
galaxies. 

The participants in the complex story of the discovery of Sgr A$^*$
are numerous: B.~Clark, D.~ Hogg, G.~ Miley, B.~ Turner, C.~ Heiles, R.~ Ekers,
D.~ Lynden-Bell, D.~ Downes, A.~ Martin, B.~ Balick, R.~ Brown, M.~ Goss,
K.Y.~ Lo, U.~ Schwarz, D.~ Rogstad and others. (Most of these are still
active astronomers.) We  present a short history of the discovery 
process (section 2) and provide some details on the naming of
Sgr A$^*$ in section 3. In section 4, we provide a short summary
of the determinations of the secular parallax of Sgr A$^*$.
Goss (2003) has also presented a summary of the discovery
of Sgr A$^*$. The process of discovery of Sgr A$^*$ was the result
of the application of a ``matched filter'' in angular resolution to the
properties of Sgr A$^*$; of course, the construction of this filter
can only be understood with an \emph{a posteriori} knowledge of the
properties of  Sgr A$^*$.

\section{The years 1965-1985}
\label{required}

In 1966, Clark \& Hogg (1966) used the newly completed 2 element Green Bank
interferometer at 11 cm to investigate the small scale structure
of a number of radio sources at 11 cm with a resolution of $\sim 10\arcsec$.
The source  Sgr A was found to have a compact feature with a flux density
of $\sim 0.3$ f.u. (Jy); with this resolution the confusion from Sgr A West
is a dominant effect. But these observers were ``close'' to the
discovery of Sgr A$^*$. We now know that a resolution
of $\sim 3\arcsec$ is required at this frequency.

The key observation that led to the discovery of the compact source at the
center of the galaxy was an observation by
Miley, Turner, Balick, \& Heiles (1970). With a baseline of 35 km
at 11 cm, these authors discovered a compact source in the \HII\/ region
W51 with a $T_b> 10^5$ K. The 42 foot telescope ,
located at the time of the Miley et al. observations at
Huntersville, West Virginia, is shown in Fig. 1. Note the limited 
tracking capability of this antenna. Although this brightness 
component in W51 has not been confirmed, the result did set off a string
of circumstances that led to the discovery of Sgr A$^*$ only four
years later.

The theoretical framework for the search for evidence of the presence of
a compact object at the galactic center was provided by Lynden-Bell (1969)
and Lynden-Bell \& Rees (1971) , who made the analogy between quasars
and the high energy phenomena at the center of the Milky Way: the latter
authors propose four tests for the possible detection of a massive
object at the center of the galaxy. The second test is :''Very Long baseline
interferometry may soon be possible with ....as weak as 0.5 f.u. to diameters
of $10^{-3}$ $\arcsec$. If so it may be possible to determine the size of any
central black hole that there may be in our galaxy. However \HII\/ may render the
central source opaque with a greater angular size.''

In the course of 1970 (January and May), Ekers \& Lynden-Bell (1971) used the newly constructed 40m 
antenna at the Owens Valley Radio Observatory (Caltech) along with one
of the original 90 foot antennas to look for
the signature of a black hole in the galactic center. At 6 cm the resolution
was $6\arcsec$ by $18\arcsec$. Ekers \& Lynden-Bell  detected fine 
scale structure in the Sgr A West  \HII\/ region.
``Although stimulated by the black hole idea our observations
are thus more simply explained in terms of young stars and giant
\HII\/ regions.'' Again if the resolution had been a factor of
about two more favorable , Sgr A$^*$ would have been detected. (Goss
has pointed out in several lectures in Australia that Ron Ekers
has mispelled his name in the acknowledgements to the
Ekers \& Lynden-Bell paper!)\   Ekers \& Lynden-Bell also performed
one of the first interferometer searches for radio recombination
lines; they used the 90 foot antenna interferometer at 6 cm. 
They searched for broad recombination lines from
Sgr A (Ekers, private communication); the  negative 
result was not reported in the Ekers \& Lynden-Bell paper. This test
had also been suggested by Lynden-Bell \& Rees (see above) to search
for the existence of a massive object at the galactic center.

The next step in the quest for compact sources at the galactic center
was the result of investigations by Downes \& Martin (1971) using the
Cambridge One Mile Telescope at 11 and 6 cm with resolutions in
RA of $\sim 11\arcsec$ and $\sim 6\arcsec$. They describe the overall
one dimensional structure (SgrA West and East) with a determination of
the spectral indices of the various components. They mention the 
presence  of structures $<10\arcsec$ in size with flux densities
$<$ $1$ Jy. Again the discovery of Sgr A$^*$ was just over the
``resolution horizon''. 

The discovery of Sgr A$^*$ did occur on 13 and 15 February 1974
by Bruce Balick and one of us ( R.L. Brown) using
the Green Bank interferometer with an 45 foot antenna
at the Huntersville West Virginia site at a distance
of about 35 km. This site was the same as the one used  in the earlier
Miley et al. (1970) observations of W51 but an improved 
antenna was now used . This antenna at the Huntersville site
is shown in Fig.2; the antenna had a wider range
of sky coverage and was operated at the dual frequencies of 11 and 3.7 cm. This
interferometer was constructed to serve as a prototype of the
Very Large Array which was under construction at the time. The
publication of ``Intense Sub-arcsecond Structure in the Galactic
Center'' was published in December, 1974 (Balick \&Brown 1974). The
resolution at 11 and 3.7 cm was $\sim 0.7\arcsec$ and $\sim 0.3\arcsec$,
respectively. With this resolution and uv coverage ( the
three simultaneous baselines from the Green Bank 3 element interferomter
and the single antenna at Huntersvile), the extended confusion from the
Sgr A West (flux density of $\sim 25$Jy and size $\sim 40\arcsec$) complex
was resolved out. Balick \& Brown write: ``The unusual structure of the
sub-arcsecond structure and its positional coincidence with the
inner 1-pc core of the galactic nucleus strongly suggests
that this structure is physically associated with the
galactic center (in fact, defines the galactic center).'' These
authors  compare the compact source with energetic nuclei of other
galaxies and even suggest that variations in the radio flux
density might be observed.

Bob Brown has unearthed a number of fascinating letters from Bruce Balick
written during the analysis period from mid March 1974 to 2 May 1974.
No copies of Bob's letters to Bruce were saved. There are
no dates on Bruce's letters. In these letters, Bruce
gives in some detail his analysis of the data and their possible
interpretations. A number of possible models for the
observed visibilities are proposed. Reading these letters today,
we are impressed with the meticulous attention to  detail in the 
interpretation of 
this difficult observation.

Here are a few amusing quotes from the letters with an emphasis
on a  fear (in retrospect somewhat unfounded) of 
competition. The time frame is toward the end of the
period March-May 1974. Bruce writes: ``Here are a few thoughts on the 45 foot Sgr A
observations. Fred Lo re-analysed some of his VLB observations of
Sgr A based on a new position I gave him and found 0.3 f.u. 
at $\lambda$ 6 cm [see
later for a description of this October 1973 observation. The detection
was only  2-3 $\sigma$]. I think his
baseline was Green Bank--Haystack [in fact, Maryland Point]. We'd 
better publish fast if we want to beat him into print. I haven't heard
from Goss or Downes. Could you call Dave Hogg [then Green Bank site
director] and ask if he's heard anything?''.

The following letter expresses some aprehension about IR competetion:
``Dave Rank and I \ [at the University of California at Santa Cruz]\ 
are going to try to detect these sources in the IR. Please keep the
positions kind of quiet, cause Becklin and co. can wipe us out
if they want to. So can Rieke. Your faithful collaborator...''

In 1975, Ekers, Goss, Schwarz, Downes, \& Rogstad (1975) combined
Westerbork (WSRT) data with Owens Valley Radio Observatory data
at 6 cm and made a 2-dimensional image of Sgr A with a resolution
of $6\arcsec \times 18\arcsec$. The image was called
the WORST image - the Westerbork Owens Valley Radio Synthesis
Telescope. In fact the beam shape looked like a sausage - ``worst ``
in Dutch. Sgr A$^*$ was just visible at the longest spacings of the
interferometer and Sgr A East, the non-thermal source that may be
a luminous supernova remnant, was clearly detected as well
as hints of the mini-spiral structure of Sgr A West , a \HII\/ region 
associated with the center of the Milky Way. 

The first VLBI attempt to detect a compact source at the
center of the Milky Way was carried out by K.Y. Lo and collaborators
in October, 1973. The observation is described by Lo (1974) in his MIT 
Phd thesis of August 1974: ``Interstellar Microwave Radiation and Early
Stellar Evolution''.  Lo was following up on the Miley et al. W51
observation of 1970, to try to confirm the detection of compact structure 
in HII regions, beyond what had been expected theoretically. The observations were at 6 cm  
between the Green Bank 140 foot (see above) and the Naval Research
Labortory 85 foot Maryland Point radio telescope. The GB-MP baseline is
mainly E-W with a length of 228 km. A source with a diameter less than 26 
mas (EW) would be unresolved and detectable down to a flux density of 
$\sim 0.1$ Jy.  As we now know, the size
of Sgr A$^*$ at 6 cm is broadened by interstellar scattering to 
an EW by NS size of $51 \times 27$ mas (Lo et al. 1998; Davies, Walsh \& 
Booth 1974). 
The orientation
of the baseline was therefore quite unfavorable for a detection.  So, 
while there were hints of a signal in the visibility amplitude, the 
detection was not definitive.  
If the baseline had been oriented in a roughly N-S direction, the source 
would have been detected.

For one of us (Goss), an amusing and somewhat embarrassing episode occurred 
in the years 1972-1974. On 2 June 1972, D.Downes and Goss (both working
at the Max Planck Institute for Radio Astronomy in Bonn, Germany) 
submitted proposal
D43 to NRAO for an observation of the galactic center with the
Green Bank 35 km radio link interferometer. The propsosal was sent
to D.Heeschen, D.Hogg and W.E.Howard.
We have been able to reconstruct all these events based on the
extensive paper archive preserved (in 2003) by Dennis Downes from  these 
pre-email days. The proposal included two positions in Sgr A and
three in Sgr B2. A few key
sentences from the proposal follow: `` In view of the increasing
interest in highly collapsed nuclear objects as probable sources
of the energy in QSO's and radio galaxies, it is of paramount importance
to pursue investigations of compact structure in Sgr A. Although the
center of the galaxy is relatively quiescent, it is so close that we
can observe details on a much finer linear scale than is possible
in external galaxies, even with VLB techniques. .... We regard this
project as an experiment which may be a useful guide to future
observations by the VLA..... These observations might be used
in future programs to investigate short-term variability in the galactic
center.'' Although these ideas were  relevant, Goss and Downes
were not able to come to the US in 1973-1974 due to problems obtaining
travel funds. Also initial  observations with the 45 foot telescope
were somewhat delayed from late 1972 to mid 1973. 
Early in 1973, Goss had moved to the
Netherlands in a visiting position at the University of Groningen
to work on WSRT projects. In addition, Downes was quite
busy with early observations with the 100 m Effelsberg telescope. With
these pressures, the urgency to complete the Downes-Goss proposal with the
Green Bank interferometer
decreased. D. Hogg had been in continual contact with
Downes and Goss about scheduling. As shown above, Balick and
Brown had an earlier NRAO proposal to observe small scale structure
(W51 type components) in \HII\/ regions and Sgr A and
Sgr B2 were included. Dave Hogg became aware of the proposal
conflict in early 1974 and wrote Downes a letter on
15 February 1974 (note the precise discovery date) proposing
several ways to resolve this conflict. However, Goss and Downes 
seemed to have lost interest at this point. Of course, the 
significant result is that Balick and Brown did discover
Sgr A$^*$ in early 1974 - in fact on 13 and 15 February.

The first successful VLBI detection of Sgr A$^*$ was made the following
year (19 May 1975) by Lo et al. (1975) using the OVRO 40 m
and the NASA Goldstone 64-m Mars antenna at 3.7 cm. K.Y. Lo had become 
a postdoctoral fellow at OVRO after his MIT thesis, but he was interested 
in following up on the tantalizing hints of detection of Sgr A* on the 
GB-MP experiment at 6 cm in 1973.  It is interesting 
to recall that after some persuasion by Lo, the observation of Sgr A* was 
added to the program of his colleagues R. Schillizzi and M. Cohen to study compact symmetric 
double radio sources.  From the California baseline, the inferred size was 
$\sim 20$ mas. 

At an URSI meeting in Boulder, probably in January 1976, after Lo 
had reported the detection of Sgr A* at 3.7cm on the OVRO-Mars baseline, 
Don Backer 
asked the interesting probing question of how one can be sure that Sgr A$^*$ 
was not a background compact radio source.  Interestingly enough, as 
indicated below, Don Backer answered his own question some years later 
when he and Dick Sramek detected the secular parallax of Sgr A$^*$ due to the 
rotation of the Sun about the Galactic Center.

In the period 10 June 1974
to 10 September 1975, Sgr A $^*$ was observed with the early MERLIN
array at 0.408, 0.96 and 1.66 GHz with baselines of
24 and 127 km. The detections at the latter two
frequencies suggested  the angular size scales as $\lambda^2$, 
originating in a turbulent electron distribution along the line of
sight (Davies, Walsh \& Booth 1974). A number of groups worked on the
subsequent VLBI observations of Sgr A$^*$ (Kellermann et al. 1977,
Lo et al. 1977, Lo et al. 1981, Lo et al. 1985, \& Lo et al. 1993). In the
1985 publication of Lo et al., this group determined for the first 
time that the scattering size of Sgr A* at 3.6 cm was asymmetrical with an
axial ratio of $\sim 0.55$, and at 1.35 cm the limit to the angular
size was 20 AU or 2.5 mas. 

The Green Bank 35 km interferometer was used to determine that the
radio spectrum of Sgr A$^*$ (Brown, Lo \& Johnston 1978) is inverted. Brown
 \& Lo (1982) carried out a ground breaking investigation of the 
variability of Sgr A$^*$ 
at 11 and 3.7 cm over a time interval of 3 years with
25 epochs (Brown \& Lo 1982); time variations were detected over
all time scales from days to years. This ground breaking project
became the basis for future detailed VLA studies of the
various scales in the time variations of Sgr A$^*$ (Zhao et al. 2001).

\section{The naming of Sgr A$^*$}

As far as we can ascertain, the only credit attributed to the naming of
Sgr A$^*$ by Brown (1982) is in the Annual Reviews article by
Melia \& Falcke (2001). The first attempt at a convenient name
of the galactic center compact source is by Reynolds \& McKee (1980)
in a paper entitled : ``The Compact Radio Source at the Galactic
Center''. This publication presents a model of  relativistic
outflows, with either a spherical or jet geometry. The fact that the
luminosity is $\sim 100$ times greater than a pulsar but
much less than other galactic nuclei was a puzzle. Reynolds \& McKee
suggest the name \emph{GCCRS}-- the galactic center compact radio source. 
This name has not survived. 

Brown \& Lo (1982) discuss the variability of Sgr A$^*$ (see above)
:'' Throughout this paper we use the name Sgr A to refer only
and specially to the compact radio source. When necessary, we distinguish
this from the more extended radio structure at the galactic
center.''

In 1982, Backer \& Sramek (1982) presented the initial results of the secular
motions of Sgr A$^*$ using the Green Bank radio link interferometer. The
motions were found to be consistent with an object at rest
in the center of the Milky Way. Backer \& Sramek propose the name: 
\emph{Sgr A(cn)} from  ``compact non-thermal'' object in the galactic center. Again this
name has had no staying power.

Eight years after the discovery, one of us (Brown) invented the name 
\emph{Sgr A$^*$}
to distinguish the compact source from the other components in the galactic
center and to emphasize the unique nature of this source. Brown (1982) proposed
a model of Sgr A$^*$ consisting of twin precessing jets with a period
of 2300 years. The model has not stood the test of time but the name
immediately was accepted. As an example, the VLBI results discussed by
Lo et al. (1985) uses the name Sgr A$^*$; the review article
by Lo (1987) also uses this nomenclature.

Bob Brown provides the following rationale for the name: `` Scratching
on a yellow pad one morning I tried a lot of possible names. When I began
thinking of the radio source as the ``exciting source'' for the cluster
of \HII\/ regions seen in the VLA maps, the name Sgr A$^*$ occurred to me by 
analogy brought to mind by my Phd dissertation, which is in atomic 
physics and where the nomenclature for excited state atoms is
He$^*$, or Fe$^*$ etc.''

\section{The motions of Sgr A$^*$}

The physical association of Sgr A$^*$ with the mass centroid of the nuclear
region of the Milky Way remained circumstantial until the observation
of the secular motion of Sgr A$^*$ by Backer \& Sramek, noted above, was 
made in 1982 with the Green Bank radio link interferomter. Even in 1983,
Martin Rees wrote to Robert L. Brown to report that at the IAU symposium
held in June 1983 in Groningen (Netherlands, IAU Symposium No 106) that Jan Oort was worried
that the lack of formaldehyde absorption toward Sgr A$^*$ could have
implied that the radio source is located nearer than the true galactic
center.

Such concerns were again soundly put to rest based on two companion papers, 
which were published in the Astrophysical Journal issue of 20 October 1999 
(Backer \& Sramek 1999; Reid et al. 1999), that 
summarize the VLA and VLBA determinations of the  motions of Sgr A$^*$.
Don Backer has pointed out to us that the initial Green Bank 35 km
radio link interferometer observation of the 1970's was inspired
by a lunch time conversation with Rick Fisher in about 1975 (see the
acknowledgement in Backer \& Sramek 1999). The secular
parallax due to the motion of the Sun around the center of the
galaxy, of course, establishes that Sgr A$^*$ is in the galactic
center, but more importantly can be used to set a lower limit
on the mass  of the black hole of a few thousand \Md\/. In addition,
a number of  constraints on galactic rotation constants can be determined.
The long range goal of the VLBA program (Reid et al 1999) is the determination
of a parallax distance to the galactic center.

\section{Summary}
The observations of the last 30 years have provided a wealth of 
information about the source Sgr A$^*$ and the environs of the
black hole at the center of the Milky Way. Many puzzles remain. We
can only imagine the contents of a possible conference on 
the center of the galaxy that might be held at the time
of the 60th celebration of the discovery of Sgr A$^*$ in 2034.
In March 2004, The National Radio Astronomy Observatory will
host a conference in Green Bank, West Virginia, to honor the discovery
of Sgr A$^*$ exactly 30 years previously and to discuss recent results
on this fascinating radio and X-ray source.

\begin{acknowledgement}
The National Radio Astronomy Observatory is a facility of the
National Science Foundation operated by Associted Universities, Inc.
under cooperative agreement.  We thank Dennis Downes, Bruce Balick,
Dave Hogg, Don Backer, Ron Ekers and Barry Clark for 
helpful comments.  Much of the work on Sgr A* by Lo was done before 
his joining the NRAO;  he appreciates the important support provided 
at the critical times in the early years by B. Burke, K. Johnston, J. 
Moran, A. Rogers, M. Cohen, R. Schillizzi, D. Backer and J. Welch.  We thank
Pat Smiley for assistance with the Green Bank interferometer photo
archive. M. Goss thanks Tony Beasley for his hospilality at the Owens 
Valley Radio Observatory 
during February, 2003, while this paper was being written.
The late J.H. Oort engendered the enthusiasm that initiated the
fascination of Goss, Ekers and Schwarz in the 1970's for galactic
center research.
\end{acknowledgement}

\begin{figure}[htb]
\begin{center}
\includegraphics[width=.80\textwidth]{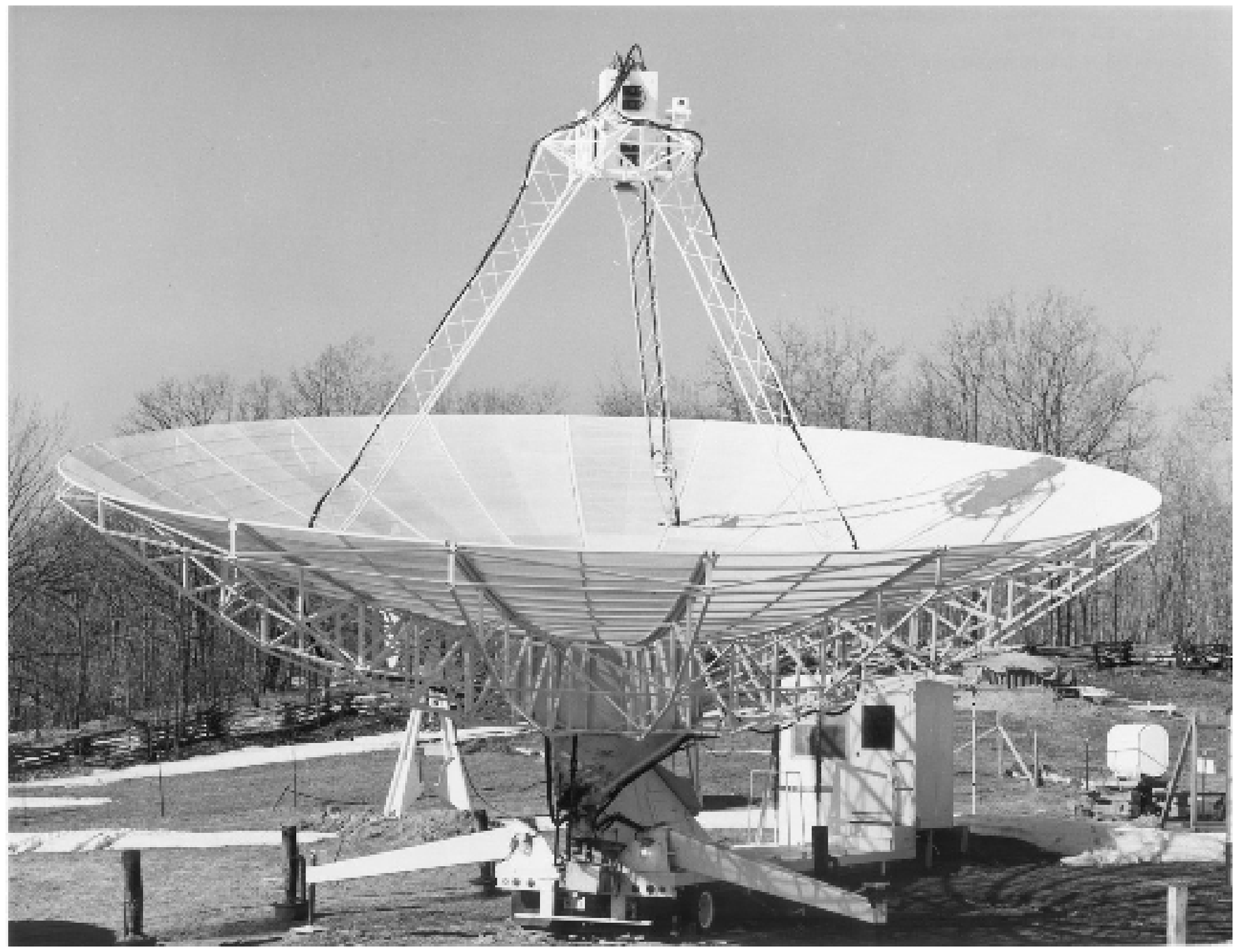}

\caption{The 42 Foot antenna.  The initial
location of this antenna was at Spencer's Ridge - 11 km NE of the
Green Bank interferometer. The operation was at 11 cm. Basart et al. 1968 and
Basart et al. 1970 describe initial observations with a two element
interferometer consisting of this antenna combined with one of the 85 Foot 
antennas
at Green Bank. The sky coverage was limited to declinations from 0 \deg to
+66 \deg and hour angles within $2^h 40^{min}$ of the meridian. 
For the observations of W51 described by Miley et al. 1970, the
antenna had moved to the Huntersville site - 35 km to the
SW of Green Bank. These observations were a partial inspiration for the
Sgr A$^*$ observations of 1974. Later the 42 Foot antenna was replaced  by the
fully steerable antenna shown in Fig.2 at the Huntersville site.
Fomalont (2000) summarizes the development of the Green Bank radio link
interferometer in the years 1966 to 1978.}
\end{center}
\label{fig:1}
\end{figure}

\begin{figure}[htb]
\begin{center}
\includegraphics[width=.9\textwidth]{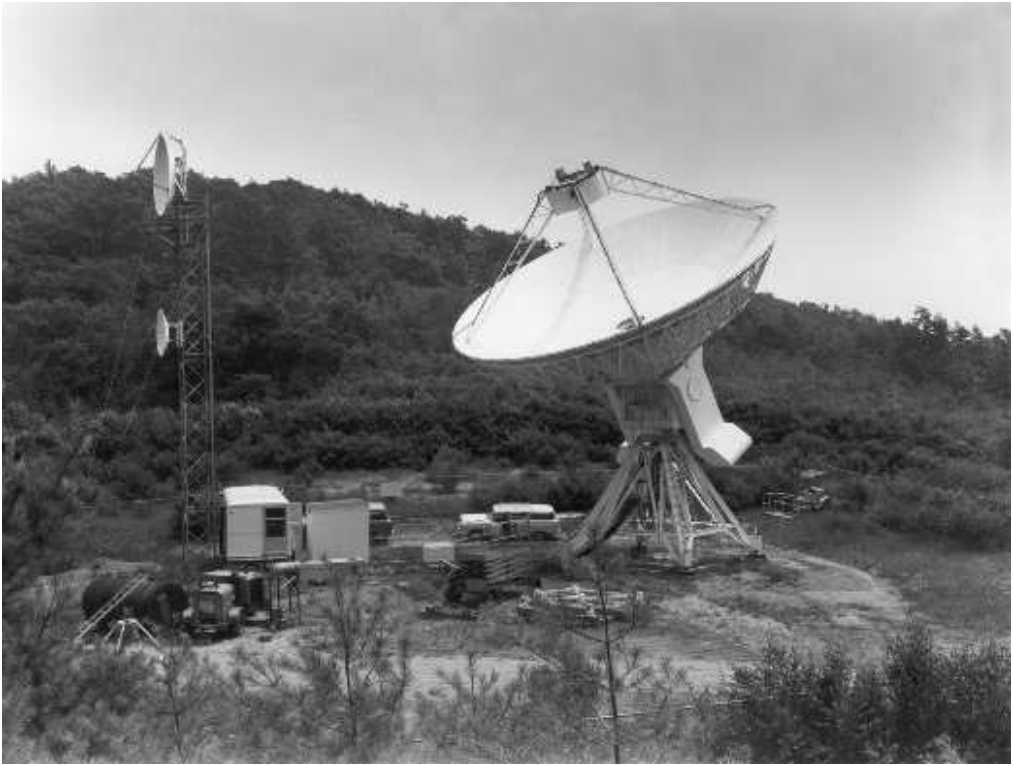}

\caption{The 45 Foot telescope used at Huntersville, West Virginia
as the 35 km outstation for the NRAO Green Bank radio link interferometer.
The discovery of Sgr A$^*$ was made in February, 1974 by Bruce Balick and
Robert L.  Brown using this instrument (Balick \& Brown 1974). The dual frequency
instrument operated at 11 and 3.7 cm and was used as a prototype
for the VLA in the planning stages during the early 1970's 
(see Fomalont 2000). The
45 Foot telescope was the fourth element of the interferometer ; correlations
were performed with all three of the 85 Foot antennas at Green Bank. 
This smaller antenna is now at the Green Bank site and has 
had an illustrious career as a component of the tracking stations for the
HALCA VLBI spacecraft. The two antennas for telemetry are pointing
to the Green Bank site to the northeast; the local oscillator signal
was transmitted by a two way link at 1.3GHz while the 18GHz link
was used for telemetry and IF transmission. During the summer there was no clear line of sight
to the main interferometer site due to leaves in intervening trees. A passive reflector
was used on a hill behind the main site to overcome this problem. Much of the
development work for the radio link was done by N.G.V. Sarma 
who was on a sabbatical
from the Tata Institute for Fundamental Research (Ooty) in India.}
\end{center}
\label{fig:2}
\end{figure}

\end{document}